# Community Evolution

Stanisław Saganowski, Piotr Bródka, Przemysław Kazienko

Wrocław University of Technology, Department of Computational Intelligence

Wrocław

Poland

stanislaw.saganowski@pwr.edu.pl, piotr.brodka@pwr.edu.pl,kazienko@pwr.edu.pl

# Synonyms

Social group evolution

Temporal communities

Evolutionary communities

Changes of social groups

# Glossary

**SN: social network**

**TSN: temporal social network**

# Definition

Evolution of a particular social community can be represented as a sequence of events (changes) following each other in the successive timeframes within the temporal social network. In other words, the evolution is described by identified group transformations from time $T_i$ to $T_{i+1}$ ($i$ is the period index).

There are several approaches to definition of possible events in the social group evolution

- Asur at al. distinguish 5 possible events that may happen to groups, i.e. they may dissolve, form, continue, merge and split [Asur 07];
- Pala et al. identify 6 distinct transformations: growth, contraction, merging, splitting, birth and death [Palla, 07];
- Bródka at al., in turn, describe 7 noticeable event types: continuing, shrinking, growing, splitting, merging, dissolving and forming [Bródka, 12].

Some other different taxonomies can be found in [Spiliopoulou, 06], [Oliveira, 10], [Takaffoli, 11] but all of them are very similar and actually complete each other. The short description of the most common changes can be found below.

**Continue [Asur, 07], continuing [Bródka, 12]** – the community continues its existence when two groups in the consecutive time windows are identical or when two groups differ by only few nodes but their sizes remain the same. Intuitively, continuation happens when two communities are so much similar that it is hard to see any significant differences.

**Contraction [Palla, 07] shrinking [Bródka, 12]** – the community shrinks/contracts when some members have left the group, making its size smaller than in the previous time window. A group can shrink either slightly, losing only few nodes, or greatly, losing most of its members.

**Growth [Palla, 07] growing [Bródka, 12]** – the community grows when some new members have joined the group, making its size bigger than in the previous time window. A group can grow slightly as well as significantly, doubling or even tripling its size.

**Split [Asur, 07], splitting [Palla, 07] splitting [Bródka, 12]** – the community splits into two or more communities in the next time window when few groups from timeframe $T_{i+1}$ consist of nodes of one group from timeframe $T_i$. Two types of splitting can be distinguished: (1) *equal*, which means the contribution of the groups in the split group is more or less the same and (2) *unequal*, if one of the groups outweighs the others and participates much higher in the split group. In the latter case, the splitting might look similar to shrinking for the biggest group.

**Merge [Asur, 07], merging [Palla, 07] merging [Bródka, 12]** – the community has been created by merging several other groups, when one group from timeframe $T_{i+1}$ consist of two or more groups from the previous timeframe $T_i$. A merge, just like the split, might be (1) *equal*, if the contribution of the groups in the merged group is almost the same, or (2) *unequal*, if one of the groups contributes into the merged group much higher than other groups. For the largest group, the merging looks quite similarly to growing in the case of unequal merging.

**Dissolve [Asur, 07], death [Palla, 07] dissolving [Bródka, 12]** happens when a community ends its life and does not occur in the next time window at all, i.e. its members have vanished or stop maintaining their relationships within the group and scattered among other groups.

**Form [Asur, 07], birth [Palla, 07] forming [Bródka, 12]** of a new community occurs when a group which has not existed in the previous time window $T_i$ comes into existence in the next time window $T_{i+1}$. In some cases, a group can be inactive even over several timeframes. Then, such sequence is treated as a dissolving of the first community and its birth again in the form of the second, new one.

The examples of all events described above are depicted in Figure 1.

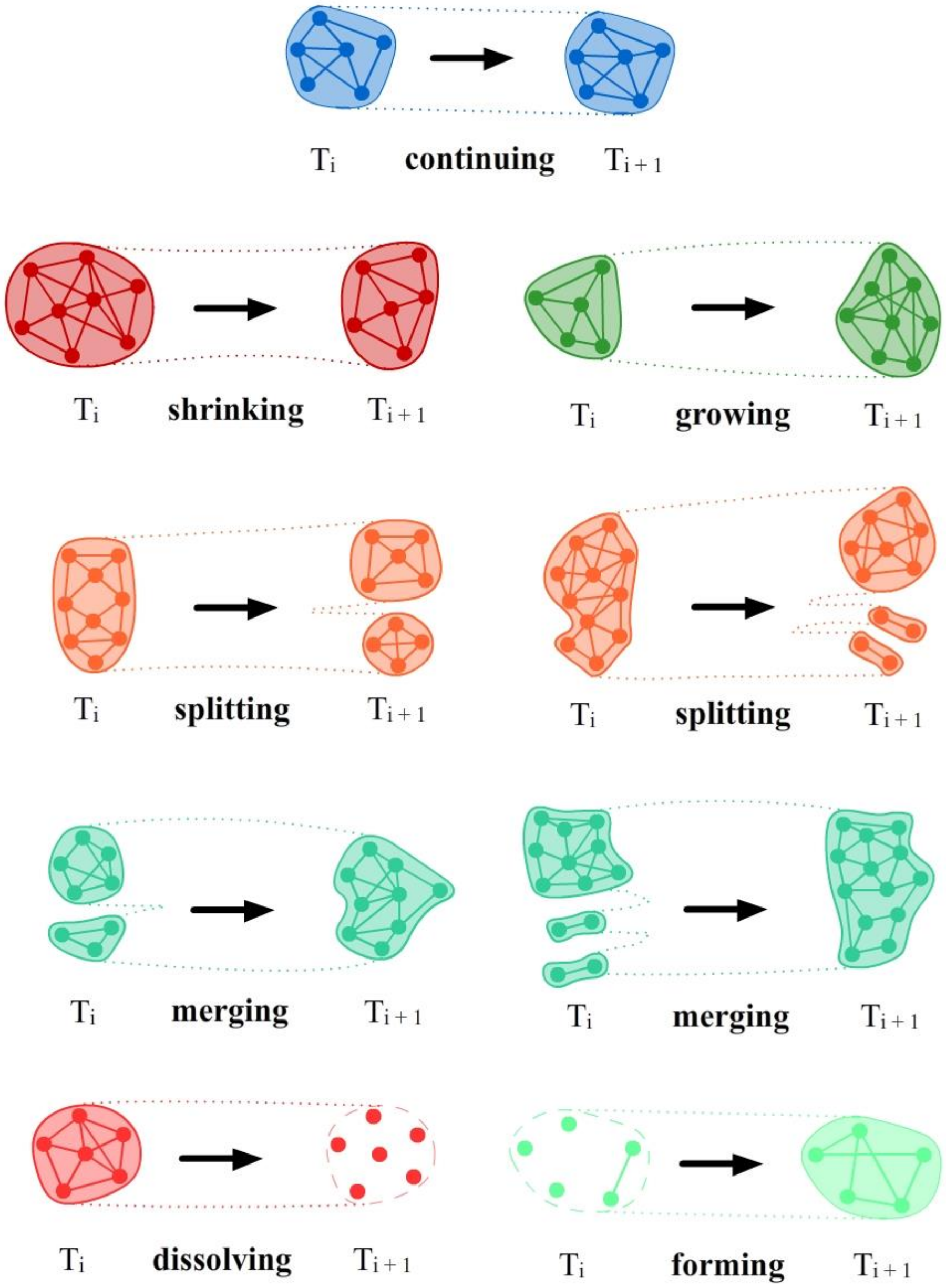

Figure 1. The events in community evolution [Bródka, 12].

The whole evolution process for a particular social community combines all changes during its lifetime to a sequence of changes - following events. A simple example of such evolution for only one group is presented in Figure 2. The community evolution is composed of seven consecutive changes, which have occurred between eight following time windows. At the beginning, group $G_1$ forms itself in $T_2$, i.e. members of $G_1$ have no relations in $T_1$ or their relations are rare. Next, the community grows in $T_3$ by gaining four new nodes. In following timeframe $T_4$, group $G_1$ splits into $G_2$ and $G_3$. By losing one node, group $G_2$ shrinks in $T_5$, while group $G_3$ remains unchanged. Then, a new group $G_4$ forms in $T_6$, while both previous communities $G_2$ and $G_3$ continue their existence. All groups merge into one community $G_5$ in timeframe $T_7$ but in the last timeframe $T_8$, this large group violently dissolves preserving only few relations between its members.

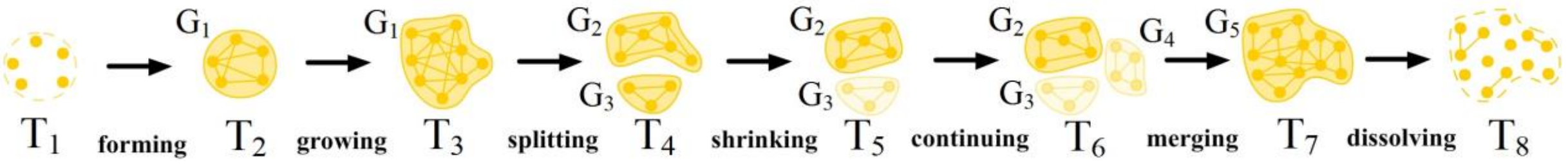

| Group in $T_1$ | Event type | Group in $T_2$ | Event type | Group in $T_3$ | Event type | Group in $T_4$ | Event type | Group in $T_5$ | Event type | Group in $T_6$ | Event type | Group in $T_7$ | Event type | Group in $T_8$ |
|---|---|---|---|---|---|---|---|---|---|---|---|---|---|---|
| No group | form | $G_1$ | grow | $G_1$ | split | $G_2$ | shrink | $G_2$ | continue | $G_2$ | merge | $G_5$ | dissolve | No group |
| | - | - | - | | | $G_3$ | continue | $G_3$ | continue | $G_3$ | | | | |
| | - | - | - | - | - | - | - | - | form | $G_4$ | | | | |

Figure 2 Changes over time for the single group.

# Introduction

The continuous interest in the social network area contributes to the fast development of this field. The new possibilities of obtaining and storing data facilitate deeper analysis of the entire social network, extracted social groups and single individuals as well. One of the most interesting research topic is the network dynamics and dynamics of social groups in particular, it means analysis of group evolution over time. It is the natural step forward after social community extraction. Having communities extracted, appropriate knowledge and methods for dynamic analysis may be applied in order to identify changes as well as to predict the future of all or some selected groups. Furthermore, knowing the most probably change of a given group some additional steps may be performed in order to change this predicted future according to specific needs. Such ability would be a powerful tool in the hands of human resource managers, personnel recruitment, marketing, telecommunication companies, etc.

To be able to describe evolution of social communities, we need to introduce the general concept of temporal social network.

First of all, a social network itself should be defined. Using a graph representation a social network *SN* is a tuple $<V,E>$, where:

*V* is a not-empty set of nodes (vertices, actors representing social entities: humans, organizations, departments etc., also called vertex or members);

*E* – is a set of directed edges (relations between actors called also arcs or connections) where a single edge is represented by a tuple $<x,y>$, $x,y \in V$, $x \neq y$ and for two edges $<x,y>$ and $<x',y'>$ if $x=x'$ then $y \neq y'$.

A temporal social network *TSN*, in turn, is a list the snapshots from the following timeframes $T_i$ (time windows, time steps). It means that each timeframe $T_i$ is in fact a single social network $SN_i(V_i,E_i)$, where $V_i$ – is a set of vertices in the *i*th timeframe and $E_i$ is a set of directed edges existing in the *i*th timeframe, as follows:

$TSN=<T_1,T_2,\ldots,T_m>$, $m$ – the total number of timeframes

$T_i=SN_i(V_i,E_i)$, $i=1,2,\ldots,m$

$E_i=<x,y>: x,y \in V_i$, $i=1,2,\ldots,m$.

An example of a temporal social network *TSN* is presented in Figure 3. It consists of five timeframes, and each timeframe is a separate social network created from data gathered within the particular interval of time. In the simplest case, one interval starts when the previous one ends, however, in some applications the intervals may overlap each other or even contain the full history of previous timeframes in the aggregated form.

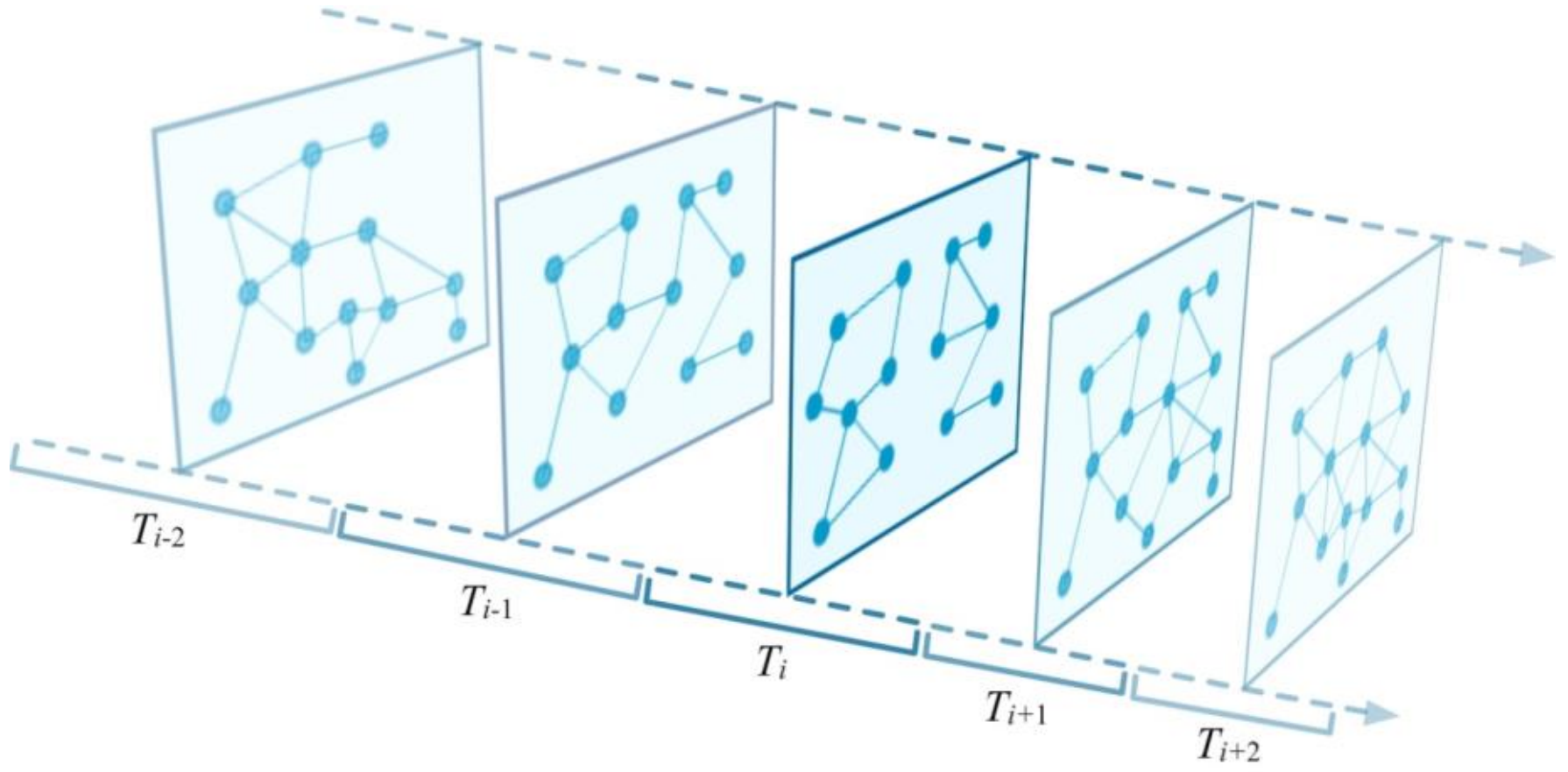

Figure. 3 A temporal social network consisting of five timeframes [Bródka, 12].

# Key Points

Several different approaches for community evolution detection can be distinguished:

1. Detection of static communities in a given timeframe and matching the separately detected communities from the following periods.
2. Detection of temporal communities also called evolutionary communities mining.
3. Evolutionary clustering, analogous to community mining.

It the first approach the data about people relationships (usually based on their activity/behaviour) is split into several timeframes forming in consequence a temporal social network. Independently, for each time window, a selected community detection method is used in order to extract social communities. Some group evolution extraction algorithms can operate on the results of one predefined group extraction algorithm like in [Palla, 07], while the other methods are independent from the grouping algorithm like [Takaffoli, 11] or [Bródka, 12]. Next, a certain similarity measure, e.g. auto-correlation function [Palla, 07], Jaccard measure [Greene, 10], inclusion measure [Bródka, 12], is utilized to match, which group from a given timeframe $T_i$ corresponds to which group in the next timeframe $T_{i+1}$. Apart from matching groups between following timeframes it is also possible to apply clustering on a graph formed by all detected groups at different timeframes [Falkowski, 06] or to calculate similarity between all the groups across all the timeframes [Tajeuna, 15]. The last step is to assign a proper change type to describe what happened between a given ($T_i$) and following snapshot ($T_{i+1}$).

The second approach also starts with creating a temporal social network, but the community detection phase is different. Instead of identification of regular, static communities for each timeframe separately, some methods to find temporal communities are applied. They detect continuous/stable social communities that last over many timeframes [Sarkar, 05], [Mucha, 10], [Kawadia, 12], [Zygmunt, 12], [Xu, 13].

Another approach is evolution of clusters, which aims to find best partition that represents the community structure at time $t$ based on partition at time $t - 1$ and information about the network at time $t$. Finding the best partition involves optimization techniques which vary across different methods. Chakrabarti et al. [Chakrabarti, 06] introduced *snapshot quality*, Sun et al. [Sun, 07] presented *encoding cost* and Lin et al. [Lin, 08] proposed *snapshot cost* to find the best partition of the network at given time. Ganti et al. [Ganti, 02] proposed a change detection framework called FOCUS, where two datasets are compared by computing a deviation measure between them. Spiliopoulou et al. [Spiliopoulou, 06] proposed an event-based framework called MONIC to model and track cluster transitions. They also introduced the concept of cluster matching to simplify the detection and evaluation of the cluster events that occurred. Oliveira et al. [Oliveira, 10] undertook dilemma of monitoring the transitions experienced by clusters over time by identifying the temporal relationships among them.

The number of methods for tracking the community evolution grows every year and it will become more and more important to develop a reliable solution to compare these methods. Granell et al. [Granell, 15] proposed a benchmark to compare static and dynamic techniques to describe group evolution. They showed that dynamic approaches are more accurate than static ones, but they evaluated only a few methods.

# Historical Background

The need to uncover and analyse community evolution derives from two important areas, namely community detection and social network evolution. Since the well-known paper by Girvan and Newman [Girvan, 02] about community structure in social networks and their method to detect them was published in 2002, dozens of new methods have appeared each year [Fortunato, 10]. At the same time, different scientists struggle to analyse, understand and model the evolution of networks [Barabasi 02], [Dorogovtsev, 03], [Kossinets, 06]. Thus, when researchers found out a little about community extraction and entire network evolution they have started to analyse the evolution of the communities themselves. Chakrabarti et al. [Chakrabarti, 06], Sun et al. [Sun, 07] and Lin et al. [Lin, 08] used partitioning to look for changes in the network over time. Kim and Han [Kim, 09] used *nano-communities* to find evolution of communities over time. Palla et al. [Palla, 07], Asur et al. [Asur, 07] and Bródka et al. [Bródka, 12] calculate similarity between groups in following timeframes in order to discover community life time. Tajeuna et al. [Tajeuna, 15] extended calculation of similarity for all the groups across all the timeframes. Xu et al. [Xu, 13] followed the contact frequency in the past between the nodes in order to track the group evolution.

# Tracking Group Evolution

One area in the social network analysis is to investigate the dynamics of a community, i.e. how a particular group changes over time. To deal with this problem several methods for tracking group evolution have been proposed. Almost all of them need as the input data the social network with communities already discovered using one of the group extraction methods. Additionally, separate methods for tracking evolution are designed to operate either on disjoint or overlapping groups and some ofthem are able to process both types. The further discussion provides the basic ideas behind the most recent methods for analysis of social group evolution and a more detailed description of three most popular methods. The summary of most representative methods can be found in Table 1.

Table 1 Methods for group evolution identification

| Name / Authors | Source | Type of communities | Type of community changes | Idea |
|---|---|---|---|---|
| Chakrabarti, Kumar, Tomkins | [Chakrabarti, 06] | disjoint | - | *snapshot quality* and *history cost* are calculated to obtain total value of partition $C_t$ at time $t$ |
| Kim, Han | [Kim, 09] | disjoint | forming, dissolving, growing, shrinking and drifting | nodes are connected to their future occurrences and neighbours with *links* creating *nano-communities*, the number and density of links determines the type of community change |
| Mucha, Richardson, Macon, Porter, Onnela | [Mucha, 10] | disjoint | - | multislice generalization of modularity obtained from the Laplacian dynamics is defined on a stacked aggregate network consisting of all snapshots |
| Takaffoli, Sangi, Fagnan, Zäıane | [Takaffoli, 11] | disjoint | split, survive, dissolve, merge, and form. | *meta community* is constructed for each series of similar groups detected by the matching algorithm in different timeframes; then, significant events are identified |
| Kawadia, Sreenivasan | [Kawadia, 12] | disjoint | - | partition distance called *estrangement* is calculated to find meaningful temporal communities |
| FacetNet / Lin, Chi, Zhu, Sundaram, Tseng | [Lin, 08] | disjoint, overlapping | - | *snapshot cost* and *history cost* are computed to obtain the appropriate partition of the data |
| GraphScope / Sun, Papadimitriou, Yu, Faloutsos | [Sun, 07] | disjoint | - | partitioning is repeated to get the smallest *encoding cost* of the graph and put it in the correct segment, jumps between segments denote changes in the graph evolution |
| Asur, Parthasarathy, Ucar | [Asur, 07] | disjoint | form, dissolve, continue, merge, split | group size and overlap between groups are calculated to assign type of the community change |
| Palla, Barabási, Vicsek | [Palla, 07] | overlapping | birth, death, growth, contraction, merge, split | groups are separately extracted from the individual timeframes, their following timeframes and the union of both to find similar communities; the type of community change is manually assigned based on the matching of groups, with their successors and unions |
| GED / Bródka, Saganowski, Kazienko | [Bródka, 12] | disjoint, overlapping | forming, dissolving, continuing, growing, shrinking, merging, splitting | *inclusion measure* is calculated to match similar communities; this measure and the group size determine the type of the community change |
| CoCE / Xu, Hu, Wang, Ma, Xiao | [Xu, 13] | disjoint | birth, death, merging, splitting, growth, contraction | *cumulative stable contact* between the nodes is computed to extract the groups and to discover the group changes |
| Tajeuna, Bouguessa, Wang | [Tajeuna, 15] | disjoint | form, dissolve, shrink, expand, split, merge, stable | matrix of similarity between the groups discovered for all the timeframes is created; based on *mutual transition* future group occurrences and types of events are defined |

## Chakrabarti et al. Method

Chakrabarti et al. presented in their method an original concept for the identifying group changes over time [Chakrabarti, 06]. Instead of extracting communities for each timeframe and matching them, the authors of the method introduced *the snapshot quality* to measure the accuracy of the partition $C_t$ in relation to the graph formation at time $t_i$. Then, the *history cost* quantifies the difference between partition $C_i$ and partition in the previous timeframe $C_{i-1}$. The total value of $C_i$ is the sum of snapshot quality and history cost at each timeframe. The most valuable partition is the one with the high snapshot quality and low history cost. To obtain $C_i$ from $C_{i-1}$, Chakrabarti et al. useed the relative weight *cp* (tuned by user) to minimize difference between snapshot quality and history cost. Chakrabarti et al. did not consider, whether their method works for overlapping groups.

## Kim and Han Method

Kim and Han in their method [Kim, 09] used *links* to connect nodes at timeframe $T_{i-1}$ with nodes at timeframe $T_i$, creating *nano-communities*. The nodes are connected to their future occurrences and to their future neighbours. Next, the authors analysed the number and density of the links to judge which case of relationship occurs for a given nano-community. Kim and Han defined the most common changes, which are: evolving, forming and dissolving. Evolving of a group can be distinguished into three different cases: growing, shrinking and drifting. Community $C_i$ grows between timeframes $T_i$ and $T_{i+1}$, if there is a group $C_{i+1}$ in the following timeframe $T_{i+1}$ containing all nodes from $C_i$. Group $C_{i+1}$ may, of course, contain additional nodes, which are not present in $C_i$. In opposite, community $C_i$ shrinks between timeframes $T_i$ and $T_{i+1}$ when there is a group $C_{i+1}$ in the next timeframe $T_{i+1}$, whose all nodes are within $C_i$. Finally, group $C_i$ is drifting between timeframes $T_i$ and $T_{i+1}$, if there is group $C_{i+1}$ in the following timeframe $T_{i+1}$, which has at least one node common with $C_i$. Kim and Han did not specify, if the method is designed for overlapping or disjoint groups, but the drifting event suggests that the method will not work correctly for overlapping groups.

## CoCE

A method by Xu et al. called CoCE [Xu, 13] aims to find the evolution of stable communities. The method first calculates the number of interactions between nodes over the time (*cumulative stable contact*, *CSC* measure). Nodes with the *CSC* greater than the threshold are joined together as the *community core*. Next, the remaining nodes are added to the *community cores*, based on the shortest distance, to form groups. Two nodes are also considered as a community.

In the following timeframes, when nodes and links are added and removed from the network, the *CSC* is calculated to decide if a group change occurred. Removing a link can cause splitting, contraction or death. Adding a link can cause birth, merging or growth.

The authors did not mention which types of the groups can be used with the method. In the experiments two data sets with disjoined groups were used.

## FacetNet

Lin et al. used evolutionary clustering to create FacetNet [Lin, 08], a framework allowing members to be a part of more than one community in a given timeframe. In contrast to Chakrabarti et al. method, Lin et al. used the *snapshot cost* and not the snapshot quality to calculate adequate of the partition to the data. Kullback-Leibler method [Kullback 51] has been used for counting snapshot cost and history cost. Based on the results of FacetNet, it is easier to follow what happens with a particular nodes, rather than what happens with a group in general. The algorithm is not assigning any events, but the user can analyse results and assign events on his own. Unfortunately, FacetNet is unable to catch forming and dissolving events.

## Tajeuna et al. Method

Tajeuna et al. with their method [Tajeuna, 15] tries to improve the methods which are looking for the match between communities only in the consecutive timeframes. In order to achieve that the groups are discovered for all the timeframes and the matrix of similarity between all of the discovered groups is created. Each community has then a vector of similarity with other communities. Groups are matched if the correlation between their representative vectors is above a threshold. This correlation is called *mutual transition*. The authors also proposed a convenient method to identify the optimal threshold value. It is unclear if the method handles overlapping groups. The experiments were conducted on three data sets with disjoined groups.

## GraphScope

Sun et al. presented parameter-free method called GraphScope [Sun, 07]. At the first step partitioning is repeated until the smallest *encoding cost* for a given graph is found. Subsequent graphs are stored in the same segment $S_i$, if the encoding cost is similar. When the examined graph $G$ has higher encoding cost than encoding cost of segment $S_i$, graph $G$ is placed to segment $S_{i+1}$. Jumps between segments marks change-points in graph evolution over time. The main goal of this method is to work with a streaming dataset, i.e. the method has to detect new communities in the network and to decide if the structure of the already existing communities should be changed in the database.

## Asur et al. Method

The method by Asur et al. is a simple and intuitive approach for investigating community evolution over time [Asur, 07]. The group size and overlap are compared for every possible pair of groups in the consecutive timeframes and events involving those groups are assigned. If none of the nodes of the community from timeframe $T_i$ occurs in the following timeframe $T_{i+1}$, the method by Asur et al. describes such case as *dissolve* of the group.

$$Dissolve\left(C_i^k\right) = 1 \text{ iff } \exists \text{ no } C_{i+1}^j \text{ such that } V_i^k \cap V_{i+1}^j > 1$$

where:

$C_i^k$ – community number $k$ in timeframe $T_i$,

$V_i^k$ – the set of the vertex (nodes) of the community number $k$ in timeframe $T_i$.

In opposite to dissolve, if none of the nodes of the community from timeframe $T_{i+1}$ was present in the previous timeframe $T_i$, the group is marked as new born.

$$Form\left(C_{i+1}^k\right) = 1 \text{ iff } \exists \text{ no } C_i^j \text{ such that } V_{i+1}^k \cap V_i^j > 1$$

A community continue its existence if an identical occurrence of the group in the consecutive timeframe $T_{i+1}$ is found.

$$Continue\left(C_i^k, C_{i+1}^j\right) = 1 \text{ iff } V_i^k = V_{i+1}^j$$

A situation when two considered communities from the timeframe $T_i$ overlap with more than $\kappa\%$ nodes of another single group in the following timeframe $T_{i+1}$, is called a merge.

$$Merge\left(C_i^k, C_i^l, \kappa\right) = 1 \text{ iff } \exists\, C_{i+1}^j \text{ such that } \frac{\left|\left(V_i^k \cup V_i^l\right) \cap V_{i+1}^j\right|}{Max\left(\left|V_i^k \cup V_i^l\right|, \left|V_{i+1}^j\right|\right)} > \kappa\%$$
$$\text{and } \left|V_i^k \cap V_{i+1}^j\right| > \frac{\left|C_i^k\right|}{2} \text{ and } \left|V_i^l \cap V_{i+1}^j\right| > \frac{\left|C_i^l\right|}{2}$$

An opposite case is marked as a split, when two groups from the following timeframe $T_{i+1}$ joint together overlap in more than $\kappa\%$ with another single group from the previous timeframe $T_i$.

$$Split\left(C_i^j, \kappa\right) = 1 \text{ iff } \exists\, C_{i+1}^k, C_{i+1}^l \text{ such that } \frac{\left|\left(V_{i+1}^k \cup V_{i+1}^l\right) \cap V_i^j\right|}{Max\left(\left|V_{i+1}^k \cup V_{i+1}^l\right|, \left|V_i^j\right|\right)} > \kappa\%$$
$$\text{and } \left|V_{i+1}^k \cap V_i^j\right| > \frac{\left|C_{i+1}^k\right|}{2} \text{ and } \left|V_{i+1}^l \cap V_i^j\right| > \frac{\left|C_{i+1}^l\right|}{2}$$

The authors of the method suggested 30% or 50% as a value for $\kappa$ threshold. Example of the events described by Asur et al. are presented in Figure 4. Communities $C_1^1$ and $C_1^2$ continue between timeframes 1 and 2, then they merge into one community $C_3^1$ in timeframe 3. In timeframe 4, community $C_3^1$ splits into three other groups $C_4^1$, $C_4^2$ and $C_4^3$, next, in timeframe 5, a new community $C_5^4$ forms and finally in timeframe 6 the biggest community $C_5^1$ dissolves.

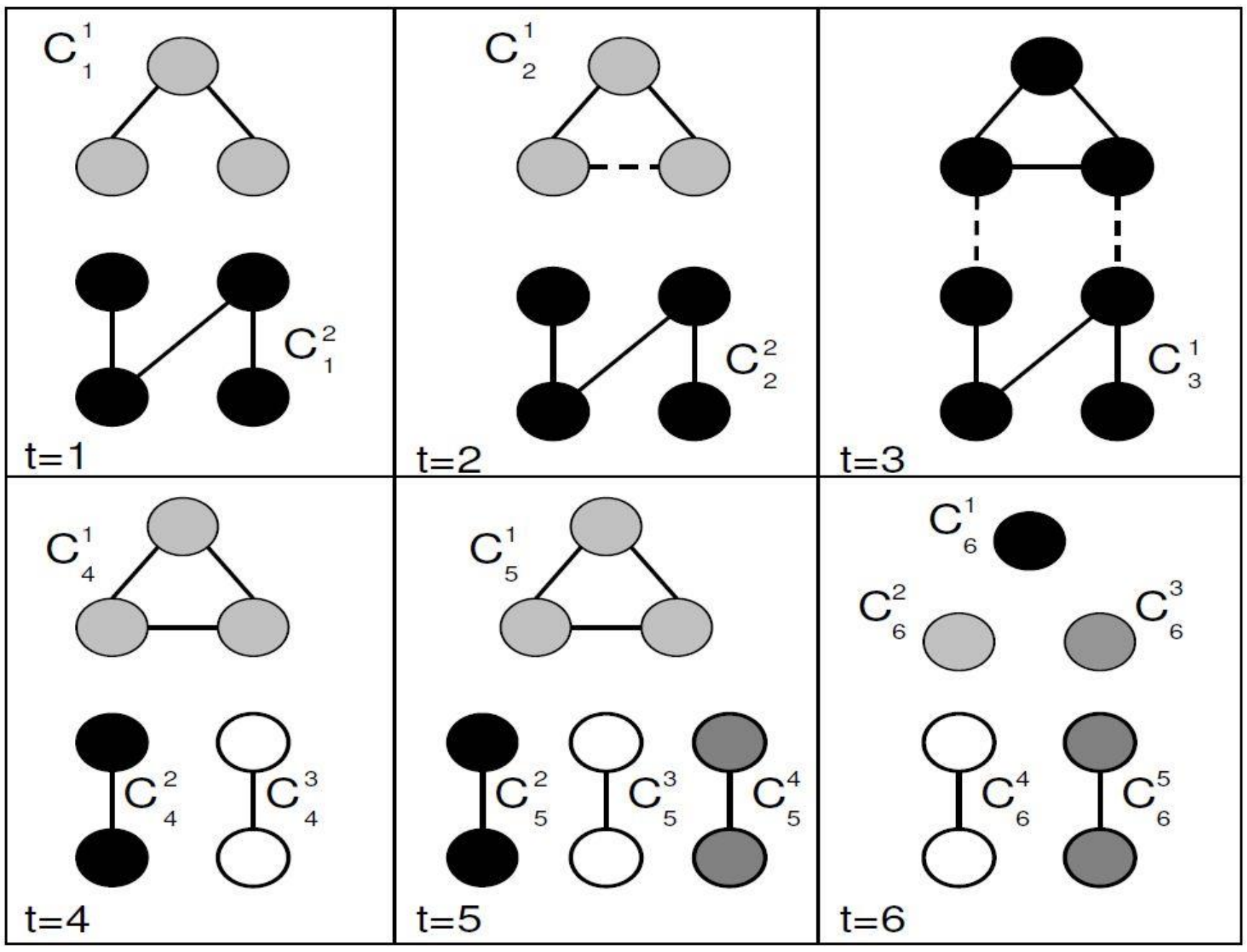

Figure 4. Possible group evolution by Asur et al. (figure from [Asur 07]).

The method proposed by Asur et al. allows also to investigate behaviour of individual members within the community lifetime. A node can appear or disappear in/from the network, and also join or leave a particular community.

Unfortunately, Asur et al. did not specify which method should be used for community detection, nor if the method works for overlapping groups.

## Palla et al. Method

Palla et al. used in their method all advantages of the clique percolation method (CPM) [Palla, 05] for tracking social group evolution [Palla, 07]. Social networks from two consecutive timeframes $T_i$ and $T_{i+1}$ are merged into a single graph $Q(T_i,T_{i+1})$ and its groups are extracted using the CPM method. Next, the communities from timeframes $T_i$ and $T_{i+1}$, which are the part of the same group from the joint graph $Q(T_i,T_{i+1})$, are considered to be matching, i.e. the community from timeframe $T_{i+1}$ is treated as an evolution of the community from timeframe $T_i$. It is quite common that more than two communities are contained in the same group from the joint graph (Figure 5b and Figure 5c). In such a case, matching is performed based on the value of their relative overlap sorted in the descending order. The overlap is calculated as follows:

$$O(C_1, C_2) = \frac{|C_1 \cap C_2|}{|C_1 \cup C_2|}$$

where:

$|C_1 \cap C_2|$ – the number of common nodes in the communities $C_1$ and $C_2$,

$|C_1 \cup C_2|$ – the number of nodes in the union of the communities $C_1$ and $C_2$.

However, the authors of the method did not explain how to choose the best match for the community, which in next timeframe $T_{i+1}$ has the highest overlap with two different groups.

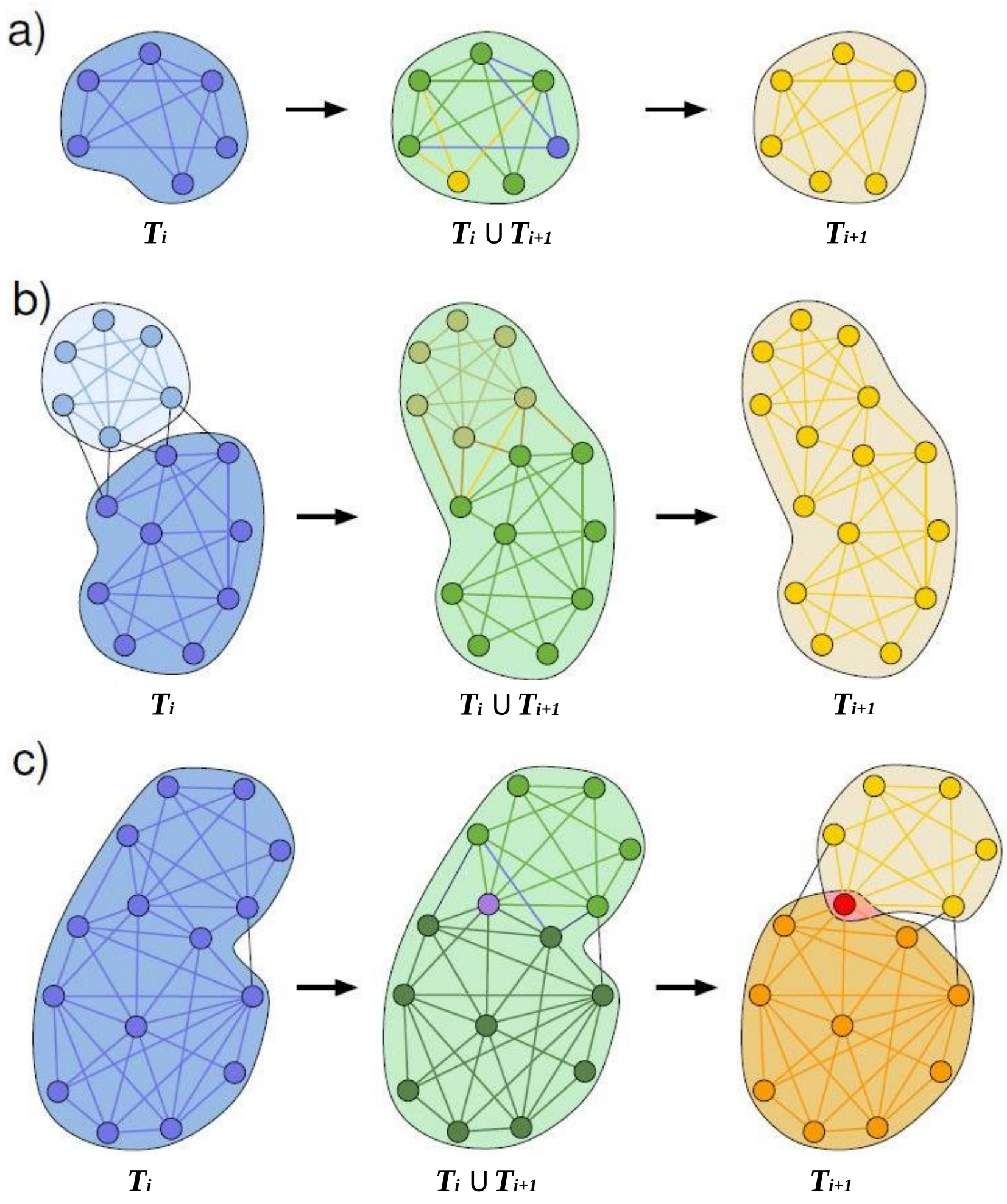

Figure 5. Most common scenarios in the group evolution by Palla et al. The groups at timeframe $T_i$ are marked with blue, the groups at timeframe $T_{i+1}$ are marked with yellow, and the groups in the joint graph are marked with green. a) a group continue its existence, b) the dark blue group swallows the light blue, c) the yellow group is detached from the orange one (figure from [Palla 07]).

Palla et al. proposed several event types between groups: growth, contraction, merge, split, birth and death, but no algorithm to identify these types has been provided. The biggest disadvantage of the method by Palla et al. is that it has to be run with CPM, no other method for community evolution can be used. Despite some lacks, the method is considered the best algorithm tracking evolution for overlapping groups.

## GED – Group Evolution Discovery

Yet another method to discover group evolution in the social network was called *GED* (Group Evolution Discovery) [Bródka, 12]. The most important component of this method is a measure called *inclusion*. This measure allows to evaluate the inclusion of one group in another. Therefore, inclusion $I(G_1,G_2)$ of group $G_1$ in group $G_2$ is calculated as follows:

$$I(G_1,G_2)=\frac{\overbrace{|G_1\cap G_2|}^{groupquantity}}{|G_1|}\cdot\underbrace{\frac{\sum_{x\in(G_1\cap G_2)}NI_{G_1}(x)}{\sum_{x\in(G_1)}NI_{G_1}(x)}}_{groupquality},$$

where $NI_{G_1}(x)$ is the value reflecting importance of node $x$ in group $G_1$.

Any metric, which indicates the member position within the community can be used as node importance measure $NI_{G_1}(x)$, e.g. centrality degree, betweenness degree, page rank, social position, etc. The second factor in the above equation would have to be adapted accordingly to the selected measure.

The *GED* method, used to discover group evolution, respects both the quantity and quality of the group members. The *quantity* is reflected by the first part of the *inclusion* measure, i.e. what portion of the members from group $G_1$ is in group $G_2$, whereas the *quality* is expressed by the second part of the *inclusion* measure, namely, what contribution of important members from group $G_1$ is in $G_2$. It provides a balance between the groups that contain many of the less important members and groups with only few but key members. A complete procedure for *GED* can be found in [Bródka, 12], whereas studies on influence of timeframe type and size are available in [Saganowski, 12].

The procedure for the Group Evolution Method (GED) is as follows:

**_GED_ – Group Evolution Discovery Method**

**Input:** Temporal social network *TSN*, in which groups are extracted by any community detection algorithm separately for each timeframe $T_i$ and any user importance measure is calculated for each group.

1. For each pair of groups <$G_1$, $G_2$> in consecutive timeframes $T_i$ and $T_{i+1}$ inclusion $I(G_1,G_2)$ for $G_1$ in $G_2$ and $I(G_2,G_1)$ for $G_2$ in $G_1$ is computed according to equations (3).
2. Based on both inclusions $I(G_1,G_2)$, $I(G_2,G_1)$ and sizes of both groups only one type of event may be identified:
    a. *Continuing*: $I(G_1,G_2) \geq \alpha$ and $I(G_2,G_1) \geq \beta$ and $|G_1| = |G_2|$
    b. *Shrinking*: $I(G_1,G_2) \geq \alpha$ and $I(G_2,G_1) \geq \beta$ and $|G_1| > |G_2|$ OR $I(G_1,G_2) < \alpha$ and $I(G_2,G_1) \geq \beta$ and $|G_1| \geq |G_2|$ OR $I(G_1,G_2) \geq \alpha$ and $I(G_2,G_1) < \beta$ and $|G_1| \geq |G_2|$ and there is only one match between $G_2$ and groups in the previous time window $T_i$
    c. *Growing*: $I(G_1,G_2) \geq \alpha$ and $I(G_2,G_1) \geq \beta$ and $|G_1|<|G_2|$ OR $I(G_1,G_2) \geq \alpha$ and $I(G_2,G_1) < \beta$ and $|G_1| \leq |G_2|$ OR $I(G_1,G_2) < \alpha$ and $I(G_2,G_1) \geq \beta$ and $|G_1| \leq |G_2|$ and there is only one match between $G_1$ and groups in the next time window $T_{i+1}$
    d. *Splitting*: $I(G_1,G_2) < \alpha$ and $I(G_2,G_1) \geq \beta$ and $|G_1| \geq |G_2|$ OR $I(G_1,G_2) \geq \alpha$ and $I(G_2,G_1) < \beta$ and $|G_1| \geq |G_2|$ and there is more than one match between $G_1$ and groups in the next time window $T_{i+1}$
    e. *Merging*: $I(G_1,G_2) \geq \alpha$ and $I(G_2,G_1) < \beta$ and $|G_1| \leq |G_2|$ OR $I(G_1,G_2) < \alpha$ and $I(G_2,G_1) \geq \beta$ and $|G_1| \leq |G_2|$ and there is more than one match between $G_2$ and groups in the previous time window $T_i$
    f. *Dissolving*: for $G_1$ in $T_i$ and each group $G_2$ in $T_{i+1}$ $I(G_1,G_2) < 10\%$ and $I(G_2,G_1) < 10\%$
    g. *Forming*: for $G_2$ in $T_{i+1}$ and each group $G_1$ in $T_i$ $I(G_1,G_2) < 10\%$ and $I(G_2,G_1) < 10\%$

The general scheme, which facilitates understanding of the event selection (identification) for the pair of groups in the *GED* method, is presented in Figure 6.

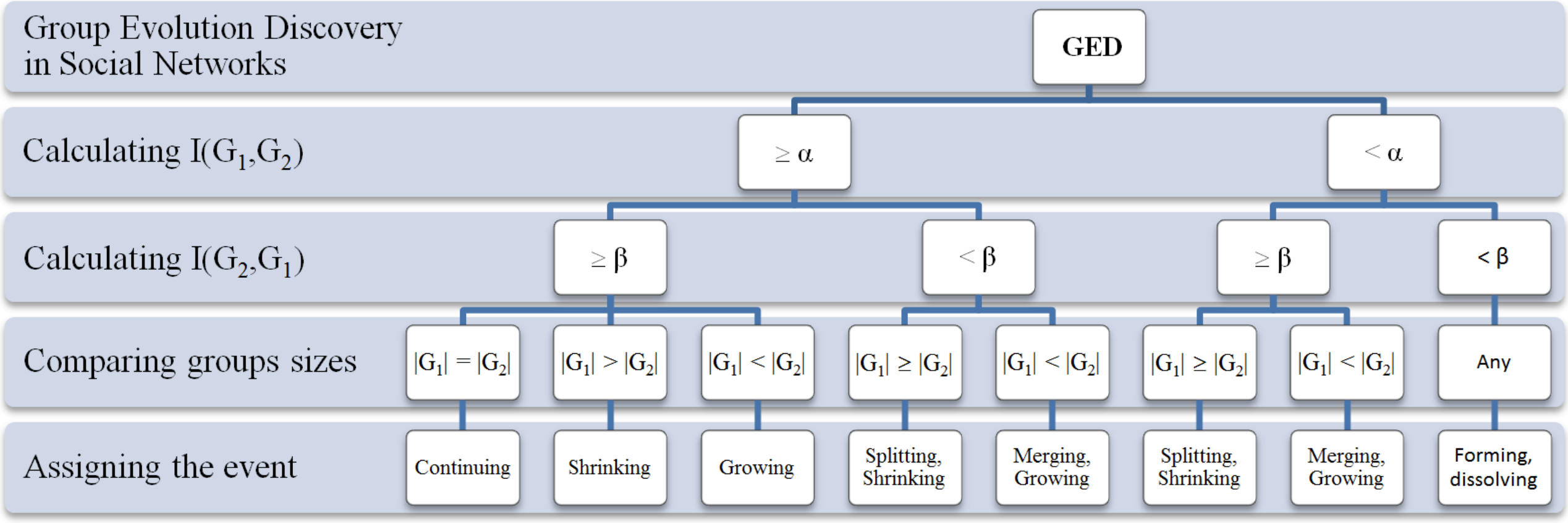

Figure 6 The decision tree for assigning the event type to a pair of groups.

Constants $\alpha$ and $\beta$ are the *GED* method parameters, which can be used to adjust the method to the particular social network and community detection method.

For example, if both $\alpha$ and $\beta$ will be set to 70% and there are two identical groups $G_2$ and $G_2$ in timeframes $T_5$ and $T_6$, respectively (see Figure 2), the inclusion measures $I(G_1,G_2)$ and $I(G_2,G_1)$ will be equal 100%. Since the size of the groups is the same *continuing* event between those groups is assigned. Another example is three groups $G_1$, $G_2$ and $G_3$ in timeframes $T_3$ and $T_4$, respectively (see Figure 2), the inclusion measures $I(G_1,G_2) = 67\%$, $I(G_2,G_1) = 100\%$, $I(G_1,G_2) = 33\%$, $I(G_2,G_1) = 100\%$. And since $G_1$ is bigger than $G_2$ and $G_3$ and there is more than one match between $G_1$ and groups in the next time window i.e. $G_2$ and $G_3$, a *splitting* event between $G_1$ and $G_2$ plus $G_1$ and $G_3$ is assigned.

# Key Applications

Detection of social group evolution is one of the crucial component of dynamic analysis of social networks. Comparison of various social groups statements enables identification of key factors that influence group evolution. It helps, for example, to answer the following question: do small groups evolve similarly as big ones?

Additionally, having changes identified some predictive models may be created in order to forecast what is most likely to happen with a certain community in the following period [Bródka, 12b], [Saganowski, 15]. Quantification of changes facilitates comparison of communities existing in various populations, e.g. among users of different services or group dynamics in different periods (this year compared to the previous one).

The possible applications span beyond typical online social networks to the analysis of group formation and evolution in face-to-face contacts networks [Atzmueller, 14] or Linux operating system network [Xiao, 17].

# Future Directions

Further research in the field of social community evolution will probably focus on extraction of useful group evolution patterns as well as analysis not only single changes between two following timeframes but long-term series of changes.

# Cross-References

Communities Discovery and Analysis in Online and Offline Social Networks, ID 00006

Social Networks, Merging and Integration, 00015

Detection, Current and Future Research Trends, ID 00027

Future Directions in Predicting the Dynamic Evolution of a Network, 00030

Temporal Networks, 00042

Evolving Social Graphs Clustering, 00047

Online Communities, 00081

Clustering Algorithms, 00138

Models for Community Dynamics, 00180

Models for Group Formation, 00181

Combining Link and Content for community detection, 00214

Connecting Communities, 00217

Extracting and Inferring communities via link analysis, 00218

Web Communities Versus Physical Communities, 00225

Social Groups in Crowd, 00255

Evolution of Social Networks, 00318

# Acknowledgements [optional]

This work was partially supported by Wrocław University of Science and Technology statutory funds and the Polish National Science Centre, decision no. 2013/09/B/ST6/02317.